# Noise enhanced performance of ratchet cellular automata

Dušan Babič[1,2] and Clemens Bechinger[1]

[1] 2. Physikalisches Institut, Universität Stuttgart, Stuttgart, Germany

[2] Faculty for Mathematics and Physics, University of Ljubljana, Ljubljana, Slovenia

We present the first experimental realization of a ratchet cellular automaton (RCA) which has been recently suggested as an alternative approach for performing logical operations with interacting (quasi) particles. Our study was performed with interacting colloidal particles which serve as a model system for other dissipative systems i.e. magnetic vortices on a superconductor or ions in dissipative optical arrays. We demonstrate that noise can enhance the efficiency of information transport in RCA and consequently enables their optimal operation at finite temperatures.

Despite tremendous achievements in the fabrication of microelectronic devices further increase in speed and circuit density will eventually require different approaches to storing and processing binary information. Rather than using voltages and currents, alternative propositions aim to encode digital data directly as positional or orientational configurations of interacting (quasi) particles or spins. In a recently proposed ratchet cellular automaton (RCA) [1] magnetic flux vortices on a patterned superconducting material were assumed for this purpose. Compared to other approaches, the novel feature of RCA is their operation far from thermal equilibrium. This is achieved by applying an external periodic drive to an assembly of confining sites, each occupied by a flux vortex. It has been shown that RCA can be applied to realize transmission lines as well as a complete logic architectures [1 2], however, fabrication of supermagnetic nano-structures with the functionality necessary for RCA construction is currently beyond the technological reach.

It is well known that static and dynamic properties of magnetic vortices are analogus to those of to classical overdamped particles [3]. Accordingly, we investigated an RCA based on colloidal particles as a model system for a magnetic vortex RCA. We found that the principle of RCA operation is robust with respect to the type of inter-particle interaction (i.e. Yukawa v.s. *1/r* interaction assumed in the simulations with magnetic vortices). The main purpose of our work, however, was to address the question how the RCA performance is affected by the thermal noise which inevitably plagues any practical device. Interestingly, we observed that noise does not necesarilly have a degrading effect but that RCA can be optimized for operation at finite temperatures. Due to the aforementioned similarity we believe that our results obtained on a colloidal RCA can be directly applied to possible realizations in magnetic vortex systems and thus may impact the design of future computing devices.

The geometry of the experimentally realized RCA corresponds to that investigated numerically by Hastings et al. [1]. Fig.1a shows a typical snapshot of 2.4 μm latex colloidal particles being trapped in several identical confining sites with 3.5 μm of mean separation in *x*-direction (schematically indicated by bright rectangular squares). The confinement is achieved by optical tweezers [4] with each site composed of three closely separated laser traps effectively forming double-well potentials oriented in *y*-direction (Fig. 1b). This reduces the particle motion to approx. one particle diameter in *y* direction (Fig. 1a). The relative intensity of the middle laser trap $I_b$ determines the double-well potential barrier. To realize a clocking mechanism crucial for RCA operation the positions of the sites have to be periodically varied in time as discussed in detail below.

Our experimental set-up is constructed around a custom built optical microscope and acousto-optical deflection system (AOD) which allows the steering and intensity modulation of an incoming laser beam [5]. The microscope consists of two coaxially placed, vertically oriented microscope objectives in a confocal configuration. The sample cell containing the colloidal suspension is placed at the focal plane between the objectives. Multiple spatiotemporally modulated optical traps are created by beam time sharing with a switching rate of up to 50 kHz. Accordingly, the light patterns can be regarded as quasi-static [6]. As light source we use an expanded 532nm laser beam which is guided through a two-axis AOD and inserted by a telecentric lens system into the entrance aperture of the upper microscope objective. The lateral addressable range of the optical tweezers in the sample plane is 150x150 $\mu m^2$, with a nanometer resolution. The intensity of each trap can be adjusted with a 12-bit resolution. Only a lateral confinement of the particles is induced by the gradient light force. In the vertical direction the negatively charged particles are pushed against the lower wall of the sample cell to a distance of approx. 500 nm where the light pressure is counter balanced by the electrostatic repulsion. The system is thus regarded as two-dimensional. The lower microscope objective together with a laser blocking optical filter is used to image the particles onto a CCD camera. All experiments are performed with a highly deionized colloidal suspension [7] in a 200µm thick fused silica cell at room temperature (stability of ± 1K).

The strength of the inter-particle electrostatic repulsion is determined by the screening length which is controlled by adjusting the salt concentration in the suspension [8] [9]. Repulsive pair-interaction leads to a zigzag shaped ground state configuration as shown on Fig. 1a. For the purpose of information processing a bit is represented by a defect where two neighboring particles are in the same state (i.e. up/up or down/down). As proposed in ref. 1, a three-phase periodic clocking mechanism applied to groups of three sites (e.g. $A_1$,$B_1$,$C_1$ on Fig.1a) is used to propagate such a defect. Assume a situation as shown in Fig.1a with a defect comprised by particles $A_3$ and $B_3$. Since the distance between sites A and B is larger than between the other sites, the electrostatic coupling of the two particles comprising the defect is weaker and the defect is stable. In the first step (clock phase I) the coupling between sites B and C is reduced by moving sites B slightly (approx. 0.7 µm) towards sites A. This makes it energetically favorable for the particle at site $B_3$ to overcome its intra-site barrier and switch its position from up to down. As a result the defect performs a step to the right. In the second step (clock phase II) the point of weakest coupling is transferred between sites C and A (by slightly moving all sites C towards sites B) inducing a further step of the defect. With the last step (clock phase III) the starting site configuration is restored by parallel movement of sites B and C back to their original positions inducing a third step of the defect. Periodic application of the described clocking sequence propagates a defect to the right along the chain. A sample video of the defect motion in colloidal RCA can be found at ref. 2.

To establish a proper operation of the RCA, the clocking period $T_c$ was typically in the range of several seconds, so the particles could follow the above described sequence. Defects were inserted at the left end of the chain by periodically ($T_{in}$) forcing the leftmost particle (input) to switch between up and down states. This was achieved with a separate independent laser trap.

A detailed analysis of the RCA operation was obtained from the recorded particle trajectories [10]. By applying suitable position thresholds we could identify the switching of individual particles within their sites and thus track the motion of defects along the chain.

Fig.2a shows the trajectories of defects traveling from the left to the right of the chain. Each step corresponds to the migration of a defect by one site as a result of the clocking sequence. Although the input particle is switched periodically and the defects travel with a constant velocity (see constant trajectory slopes in Fig.2a), they are unevenly spaced within the chain. The same phenomenon is also reflected in the bimodal structure of the waiting time distribution ($P(T_w)$ in the inset of Fig.2a) for the rightmost particle (output). This is because a defect can only be inserted into the chain during clock phase I (as defined above). Accordingly, a single peak is only observed if clocking and input particle switching periods are commensurate which is not the case in Fig.2a ($T_c$ = 4s and $T_{in}$ = 13s). It follows from the same argument that the upper frequency limit for the defect insertion is determined by $T_c$.

Fig.2a also shows that step heights of the trajectories slightly differ thus causing the peaks in the waiting time distribution of the output particle to be broadened. This is due to the thermal motion which causes particles not to follow the external clocking in a completely deterministic way. The same is observed in the simulations at finite temperatures. Nevertheless, the overall propagation of defects in this regime is deterministic as they are propagated by three steps per one clock period.

So far, particle configuration changes were caused by the externally applied clocking sequence and input particle switching (deterministic regime). It should be noted, however, that this kind of operation can only be realized for a rather narrow range of experimental parameters. The parameters governing the behavior of RCA are the temperature, the mean inter-particle coupling and the intra-site barrier height. Depending on them the defect propagation can become uneven (non-deterministic regime) as shown in Fig.2b. When there is more than one defect present in the chain this may lead to crossing and consequent annihilation of defect pairs (Fig. 2b). All these effects are reflected in a significant broadening of the output particle waiting time distribution (inset of Fig. 2b).

To investigate the non-deterministic regime of a RCA operation in more detail we analyzed the defect *propagation efficiency* $\eta$, i.e. the ratio of the number of steps a defect needs to traverse a certain distance in the entirely deterministic regime to the number of the steps actually required (averaged over many defects). The plot of this quantity vs. $I_b$ (inversely proportional to the barrier height) is shown in Fig.3 for different inter-particle couplings (multiple curves). For large barriers ($I_b < 0.5$) the particles are strongly locked to the zigzag ground state and the propagation efficiency is low. On decreasing the barrier the defect propagation becomes more efficient until reaching a maximum at some optimal value. Further decrease of the barrier ($I_b > 0.8$) degrades the propagation efficiency which is due to the thermal noise overwhelming the deterministic particle switching dictated by the clocking mechanism. Under this conditions, besides defect annihilation, we also observe spontaneous creation of defect pairs (not shown).

A similar, i.e. non monotonic behavior as in Fig.3, is also observed if we analyze the switching of individual particles (signal) at a given site in the presence of a periodic input. For this purpose the motion of each particle is first mapped to a two level signal by applying the same thresholds as used for defect tracking. This two level signal is then spectrally analyzed and its spectral density is obtained by integration over all spectral peaks.

Fig.4 shows the spectral density of the signal at different sites along the ratchet chain in the non-deterministic regime. Signal decays with the initial slope $\beta$ depending on $I_b$ (i.e. barrier height) as well as the inter-particle coupling (data shown only for one coupling strength). Again, we observe a non-uniform behavior, indicating that the ratchet transmits the signal optimally only for a certain range of barrier heights (inset of Fig. 4).

It is important to mention that the defect propagation efficiency and the decay rate of the signal are not describing identical properties of the RCA. The defect propagation efficiency is a measure of the average speed of an individual defect and depends only on the properties inherent to the ratchet. The decay rate of the signal, on the other hand, is determined by the loss of the coherent motion of defects and thus by their stalling, annihilation and creation. These effects are more likely to affect higher frequency signals where defects are more densly packed within the transmission line.

Besides inter-particle coupling strength and the intra-site potential barrier height, several other parameters determine whether the RCA is operated in the deterministic or non-deterministic regime. These are the thermal noise level, the overall confinement of the particles in their confining sites, the clock speed, the geometrical arrangement etc. During our experiments these parameters were kept constant. Thermal noise, specifically, is determined by the temperature and is an intrinsic ingredient of a colloidal system. Its level could be varied but only in the narrow range of temperatures accessible with a colloidal suspension. Nevertheless, we expect that in the complementary situation with the barrier height and the coupling strength fixed but with the variation of the noise level, the RCA will show a similar non-monotonic behavior thus performing best at some optimal noise level. In this complementary picture a link to phenomena resembling stochastic resonance (SR) [11] is established. Many examples of non-equilibrium systems exhibiting SR features have been identified and extensively studied. Nevertheless, noise enhanced performance in complex information processing systems [12] [13] was beyond the grasp of the theory. Our practical realization of an RCA, however, provides an experimental model amenable to detailed theoretical investigations.

In summary, we have constructed an RCA in a colloidal system. Our results demonstrate that the performance of the RCA crucially depends on the inter-particle coupling strength and the intra-site potential barrier height. With these parameters the ratchet can be optimized for the operation at finite temperatures. Due to the demonstrated robustness of the principle of operation and well established analogy of colloidal to other (quasi) particle systems these findings may be useful for the design of RCA devices based on other principles e.g. magnetic vortices or ions trapped in dissipative optical light arrays[14].

The authors wish to acknowledge helpful discussions with Charles Reichhardt, Cynthia Olson-Reichhardt, and Igor Poberaj. This work is supported by the Deutsche Forschungsgemeinschaft, grant BE 1788 4/1.

**Figure 1**

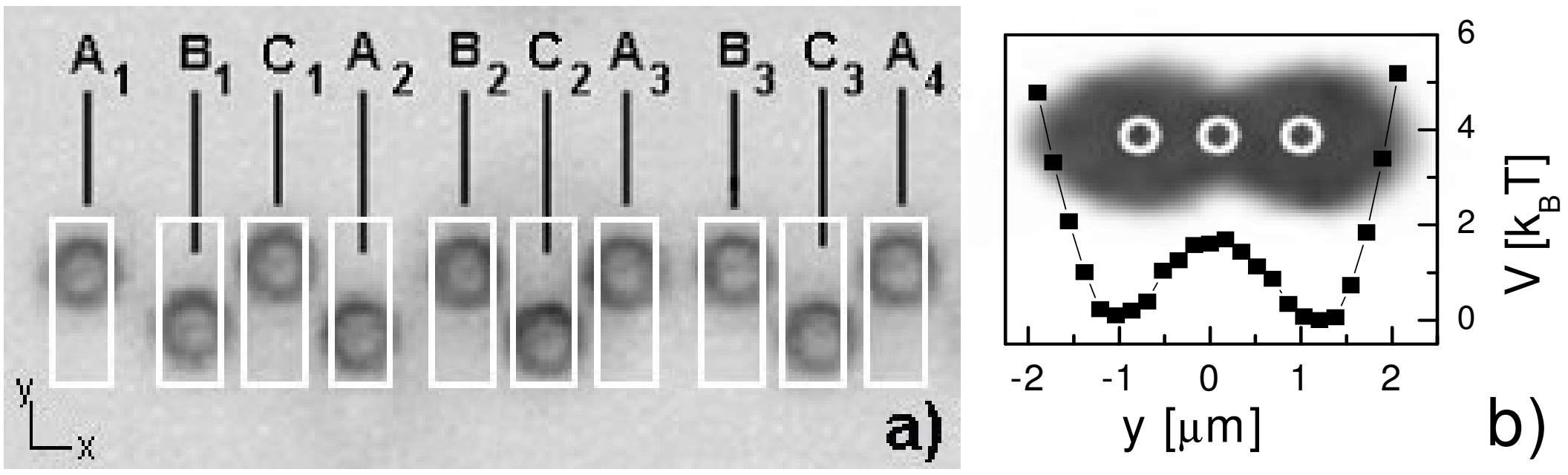

Figure 1 a) Part of a RCA transmission line. A three phase modulation of site positions propagates the defect A3/B3 to the right of the chain (see main text). b) A single confining site with the laser trap positions denoted by bright circles. The plot shows a cross-section of the light potential. The barrier height is determined by the intensity of the middle trap relative to the side ones ($I_b$).

**Figure 2**

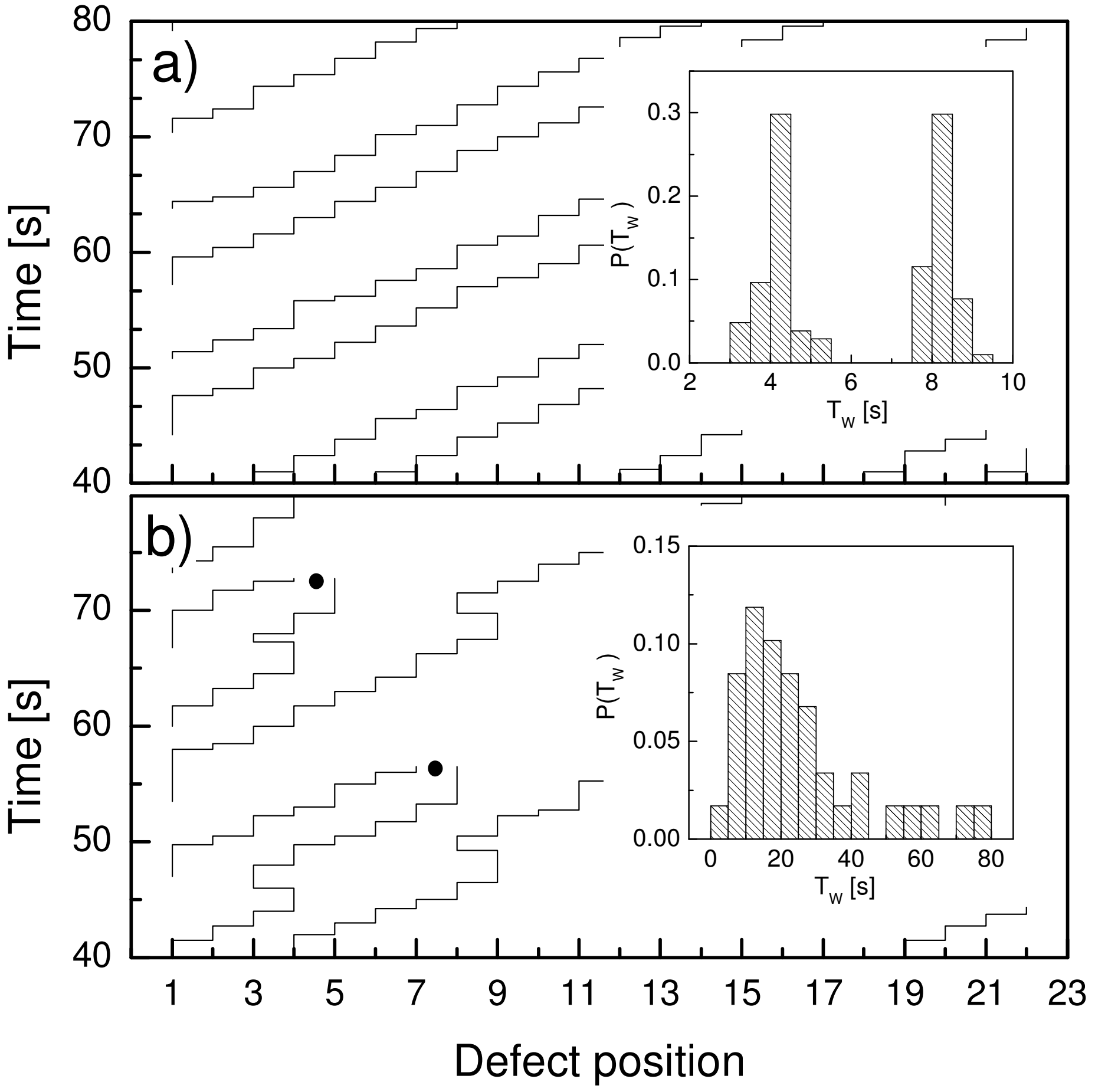

Figure 2 Defect trajectories. a) Deterministic mode: each defects makes three steps per clock period. b) Non-deterministic mode: the defect propagation is randomly stalled. Dots indicate annihilation events where two defects crossed. The insets show the waiting time distribution $P(T_w)$ for the rightmost particle (output).

**Figure 3**

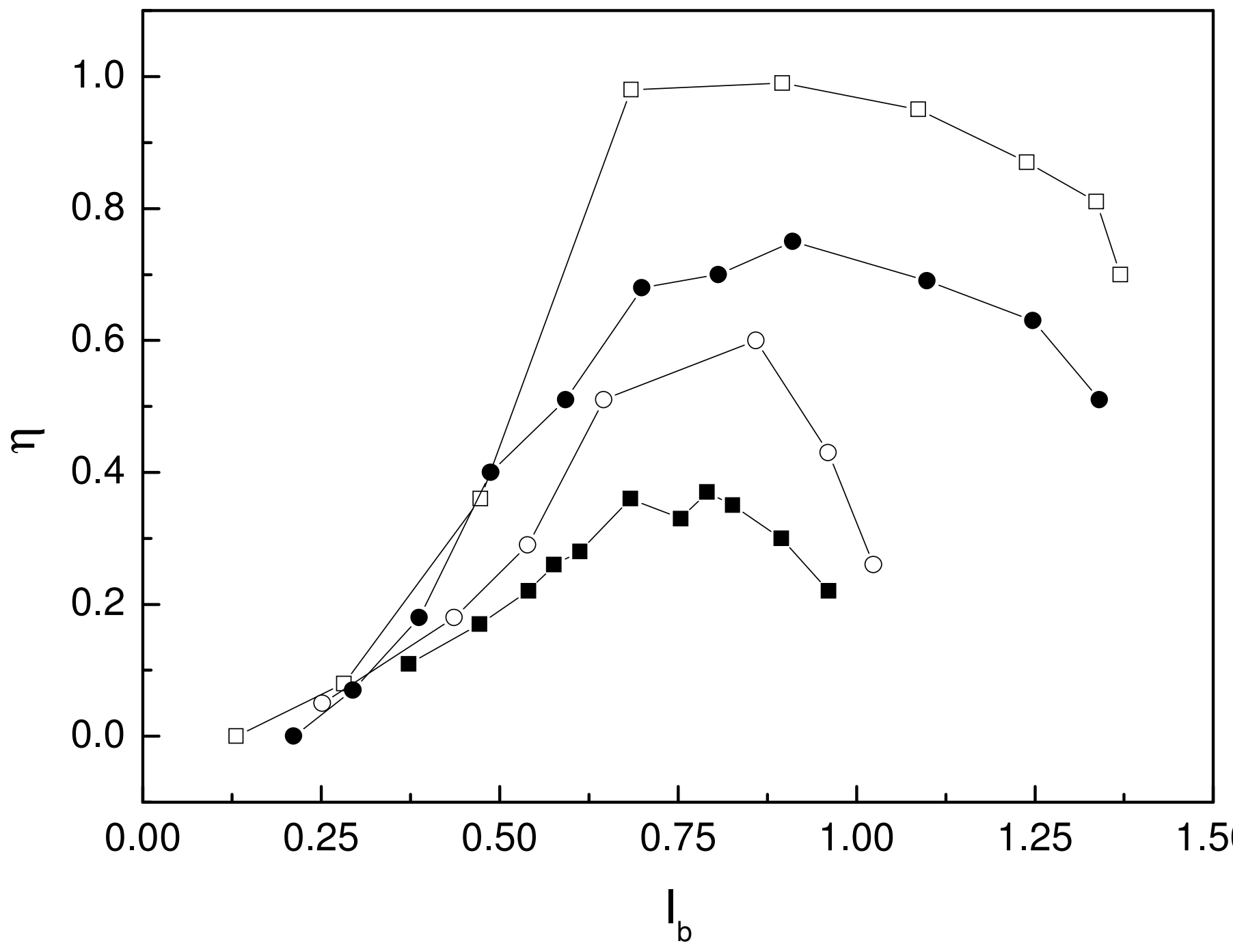

Figure3 Defect propagation efficiency $\eta$ vs. the relative middle trap intensity $I_b$ for increasing strength of the inter-particle coupling in the order: ■, ○, □, ●. Optimal performance is achieved for limited range of $I_b$ (or barrier heights), reaching a completely deterministic regime ($\eta$ = 1) only for a limited range of coupling strengths.

**Figure 4**

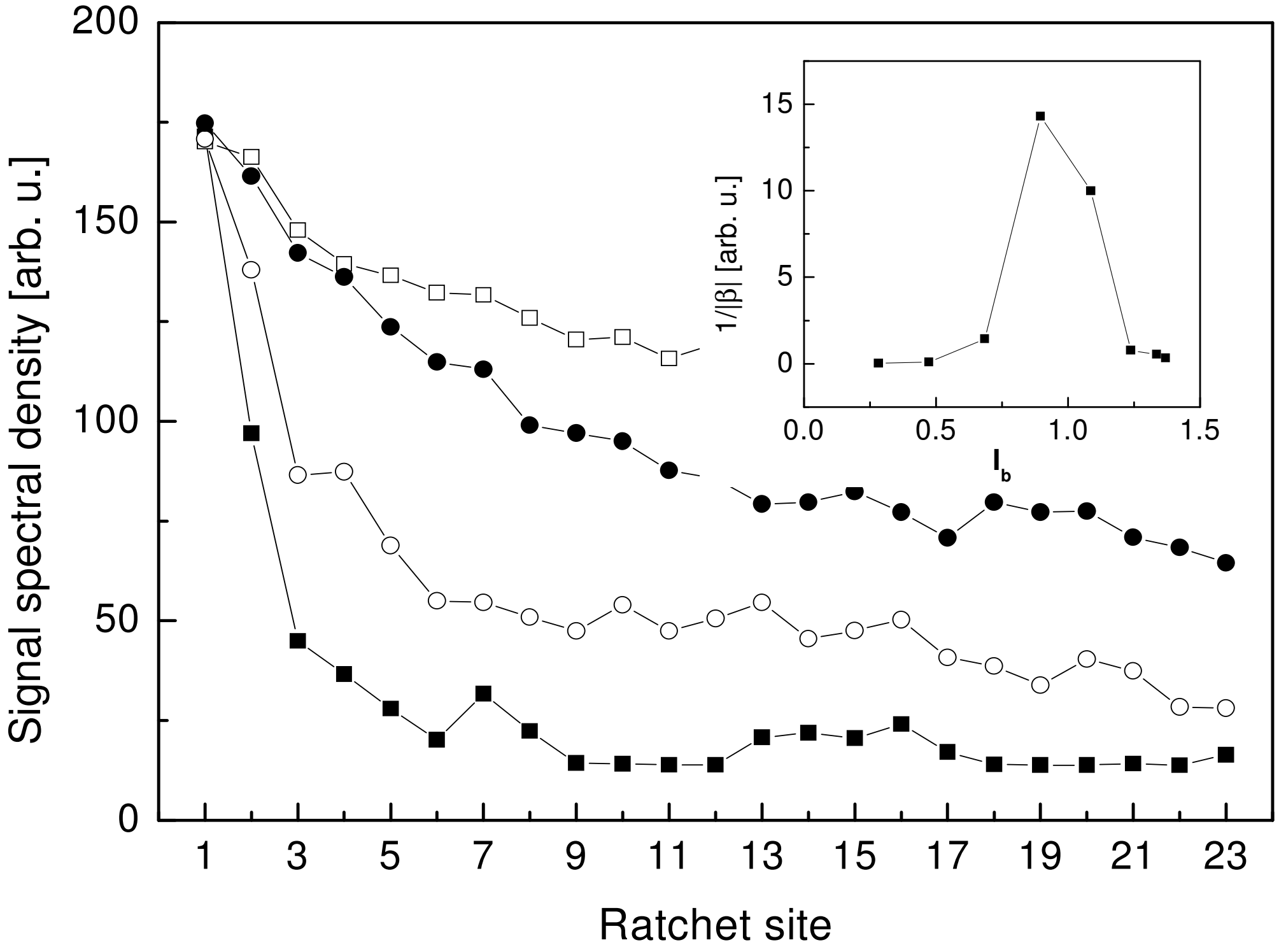

Figure 4 Signal decay along the RCA transmission line. In the non-deterministic mode the rate of signal decay (initial slope $\beta$) depends on the barrier height (decreased in the order: ■, ○, □, ● - not all values shown). Inset: optimal signal transmission is obtained for a narrow range of barrier heights (data shown for only one inter-particle coupling strength).